\documentclass[prd,showpacs]{revtex4}

\usepackage{amsmath}
\usepackage{amsthm}
\usepackage{amssymb}
\usepackage{amscd}
\usepackage[british]{babel}
\usepackage{graphicx}
\usepackage{psfrag}
\usepackage{epsfig}
\usepackage{rotating}
\usepackage{times}

\theoremstyle{plain}

\newcommand{\boxend}{\flushright{$\Box$}}

\newcommand{\R}{{\mathbb R}}               

\renewcommand{\Re}{\mbox{\rm Re}}
\renewcommand{\Im}{\mbox{\rm Im}}

\newcommand{\w}{\omega}
\newcommand{\f}{\frac}

\renewcommand{\tilde}{\widetilde}

\begin{document}

\title{Physically Sound Hamiltonian Formulation of the Dynamical Casimir Effect}

\author{Jaume Haro$^{1,}$\footnote{E-mail: jaime.haro@upc.edu} and
Emilio Elizalde$^{2,}$\footnote{E-mail: elizalde@ieec.fcr.es,
elizalde@math.mit.edu}}

\affiliation{$^1$Departament de Matem\`atica Aplicada I, Universitat
Polit\`ecnica de Catalunya, Diagonal 647, 08028 Barcelona, Spain \\
$^2$Instituto de Ciencias del Espacio (CSIC) and Institut
d'Estudis Espacials de Catalunya (IEEC/CSIC)\\ Universitat Aut\`{o}noma de Barcelona, Torre C5-Parell-2a planta, 08193 Bellaterra (Barcelona) Spain}

\thispagestyle{empty}

\begin{abstract}
Recently [J.~Haro and E.~Elizalde,~Phys.~Rev.~Lett.~{\bf 97},~130401
(2006)], a Hamiltonian formulation has been introduced in order to
address some longstanding severe problems associated with the physical
description of the dynamical Casimir effect at all times while the
mirrors are moving. Here we present the complete calculation providing
precise details, in particular, of the regularization procedure, which
is decisive for the correct derivation of physically meaningful quantities.
A basic difference when comparing with the results previously
obtained by other authors is the fact that the motion
force derived in our approach contains a reactive term
---proportional to the mirrors' acceleration. This is of the essence
in order to obtain particles with a positive energy all the
time during the oscillation of the mirrors ---while always satisfying
the energy conservation law. A careful analysis of the
interrelations among the different results previously obtained in
the literature is then carried out. For simplicity, the specific case of a neutral
scalar field in one dimension, with one or two partially
transmitting mirrors (a fundamental proviso for the regularization
issue) is considered in more detail, but our general method is shown to be generalizable,
without essential problems (Sect. 2 of this paper), to fields of any kind in two and
higher dimensions.
\end{abstract}


\maketitle





\section{Introduction}

Moving mirrors modify the structure of the quantum vacuum, what
manifests in the creation and annihilation of particles. Once the
mirrors return to rest, a number of the produced particles will
generically still remain, which can be interpreted as radiated
particles. This flux has been calculated in the past in several
situations by using different methods, as averaging over fast
oscillations \cite{dk96,jjps97}, by multiple scale analysis
\cite{cdm02}, with the rotating wave approximation \cite{sps02},
with numerical techniques \cite{ru05}, and others \cite{bmoo}. Here
we will be interested in the production of the particles and in their
possible energy values, all the time while the mirrors are in movement. This is
in no way a simple issue and a number of problems have recurrently appeared in the
literature when trying to deal with it. To start with, it is far
from clear which is the appropriate regularization to be used.
Different authors tend to use different regularizations, forgetting
sometimes about the need to carry out a proper (physical) renormalization
procedure, in order to obtain actually meaningful quantities. Thus,
it turns out that
ordinarily, in the case of a single, perfectly reflecting mirror,
the number of produced particles as well as their energies diverge all the time
while the mirrors move. Several  prescriptions have
been used in order to obtain a well-defined energy, however, for
some trajectories this finite energy is {\it not} a positive quantity and
cannot be identified with the energy of the produced particles (see
e.g. \cite{fd76}).

Our approach relies on two very basic ingredients \cite{he2}. First, proper use of a
Hamiltonian method and, second, the introduction of partially
transmitting mirrors, which become transparent to very high
frequencies. We will here prove in this way, both that
the number of created particles remains finite and also that their
energies are always positive, for the whole trajectories corresponding to
the mirrors' displacement. We will also calculate from first
principles the radiation-reaction force that acts on the mirrors
owing to the emission and absorption of the particles, and which is
related with the field's energy through the ordinary energy conservation law.
This carries along, as a consequence, that the energy of the field at any
time $t$ equals, with the opposite sign, the work performed by
the reaction force up to this time $t$ \cite{fv82,be94}. Such force
is usually split into two parts \cite{e93,bc95}: a dissipative force
whose work equals minus the energy of the particles that remain
\cite{fv82}, and a reactive force, which vanishes when the mirrors return
to rest. We will also prove below that the radiation-reaction force
calculated from the Hamiltonian approach for partially transmitting
mirrors satisfies, at all time during the mirrors' oscillation, the
energy conservation law and can naturally account for the creation
of positive energy particles. Also, the dissipative part obtained within
our procedure agrees with the one calculated by other methods, as using the
Heisenberg picture or other effective Hamiltonians. Note, however,
that those methods have traditionally encountered problems with the
reactive part, which in general yields a non-positive energy that
cannot be considered as that of the particles created at any
specific $t$.

The organization of the paper is as follows. In Sect.~2 we
introduce first the canonical formulation underlying the whole procedure. In
particular, we give the explicit expressions for the Hamiltonian and
the corresponding energy. We do this by considering the Hamiltonian
method for a neutral Klein-Gordon field in a cavity, generically
in $(n+1)-$ [although we will mainly work in $(3+1)-$] dimensional space-time,
with boundaries moving at a certain speed $v \ll c$. In
Sect.~3 we deal with the case of a single partially transmitting
mirror. We formulate the quantum theory based in the Hamiltonian
approach and, successively, the quantum theory in the Heisenberg
picture. We finish that section with a detailed comparison with
early known results, e.g. those obtained with the method of Jaekel
and Reynaud, and with the method of Barton and Calogeracos. In
Sect.~4 we study the more difficult case of two partially
transmitting mirrors. In this situation, the part of the Hamiltonian
that describes the interaction between the field and the mirrors
depends on $\epsilon$ what, in general, renders it quite difficult to
describes this part. For that reason, the reactive part of the
motion force can seldom be calculated. In any case, we prove here,
in particular, that our dissipative part of the motion force exactly
coincides with the dissipative force obtained by Jaekel and Reynaud
\cite{jr9236}. Moreover,
following the Hamiltonian approach we show that the problem of the
negative energy that appears in the Davis-Fulling model can be
resolved if partially transmitting mirrors are considered, what is in our view a very
physical approach to the renormalization issue. The last section of the paper
is devoted to a final discussion and conclusions.

\section{Canonical formulation of the problem}

We consider a neutral massless scalar field in a cavity, $\Omega_t$,
and assume that the cavity boundary is at rest for all times $t\leq
0$ and returns to its initial position at time $t=T$, to remain there
for a while. Its velocity will be given in terms of $c$, so that we
will work with the dimensionless quantity $\epsilon =v/c\ll 1$. In a
practical situation, as the one featured in Ref.~\cite{kbo1}, this
turns out to be of order $10^{-8}$ (more about that later).

The Lagrangian density of the field is
\begin{eqnarray}
{\mathcal L}(t,{\bf x})=\f{1}{2}\left[(\partial_t\phi)^2
-|\nabla_{\bf x}\phi|^2\right],\quad \forall {\bf
x}\in\Omega_t\subset \R^n, \ \, \forall t\in \R,
\end{eqnarray}
If we use the canonical conjugated momentum
\begin{eqnarray}
 {\xi}(t,{\bf x})\equiv\f{\partial {\mathcal L}}
{\partial( {\partial_t\phi})}= {\partial_t\phi}(t,{\bf x}),
\label{(2)}
\end{eqnarray}
the energy density of the field is given  by the expression
\begin{eqnarray}&&
{{\mathcal E}}(t,{\bf x})\equiv
{\xi}{\partial_t\phi}-{\mathcal L}(t,{\bf x})
=\f{1}{2}
\left({\xi}^2
+|\nabla_{\bf x}\phi|^2\right),
\end{eqnarray}
and the energy itself is \begin{eqnarray}
E(t;\epsilon)\equiv\int_{\Omega_t}d^n{\bf x}\
{{\mathcal E}}(t,{\bf x}).
\end{eqnarray}

\subsection{Hamiltonian and energy}

It is a well-known fact that the energy density does not generically coincide
with the Hamiltonian density \cite{rt85}-\cite{sps98}. Here, to obtain the
Hamiltonian  density of the field
we follow the method discussed in Refs.~\cite{js96} and
\cite{haro1}.
First, we transform the moving boundary into a fixed one by
performing a (non-conformal) change of coordinates
\begin{eqnarray}
{\mathcal R}:(\bar{t},{\bf y})\rightarrow
(t(\bar{t},{\bf y}), {\bf x}(\bar{t},{\bf y}))=(\bar{t}, {\bf R}
(\bar{t},{\bf y})),
\end{eqnarray}
that transform the domain $\Omega_t$ into a domain $\tilde\Omega$
which is independent of time. Making use of the coordinates
$(\bar{t},{\bf y})$, the action of the system behaves as
\begin{eqnarray}
S=\int_{\R}\int_{\tilde{\Omega}} d^n{\bf y}\ d\bar{t} \
\tilde{\mathcal L}(\bar{t},{\bf y}),
\end{eqnarray}
with $\tilde{\mathcal L}(\bar{t},{\bf y}) \equiv J{\mathcal
L}({\mathcal R}(\bar{t},{\bf y}))$, where we have introduced the
Jacobian $J$ of the coordinate change, defined by $d^n{\bf x}\equiv
Jd^n{\bf y}$.

Let us now consider the function $\tilde{\phi}$ given as
 $\tilde{\phi}(\bar{t},{\bf y})\equiv\sqrt{J}
{\phi}({\mathcal R}(\bar{t},{\bf y}))$. Then, the canonical
conjugated momentum is
\begin{eqnarray}&&
 \tilde{{\xi}}(\bar{t},{\bf y})\equiv\f{\partial \tilde{\mathcal L}}
{\partial( \partial_{\bar{t}}\tilde{{\phi}})}=
\partial_{\bar{t}}\tilde{{\phi}}-\f{1}{2}\tilde{{\phi}}\partial_{\bar{t}}
(\ln J)
+<{\bf y}_t,{\bf\nabla}_{\bf y}\tilde{{\phi}}-
\f{1}{2}\tilde{{\phi}}{\bf\nabla}_{\bf y}(\ln J)>
\nonumber
\\&& \hspace{1.1cm}=\sqrt{J}\partial_t\phi({\mathcal R}(\bar{t},{\bf
y})), \label{(6)}
\end{eqnarray}
and, from here, the  Hamiltonian density is obtained as
\begin{eqnarray}&&
\tilde{{\mathcal H}}(\bar{t},{\bf y})\equiv
\tilde{{\xi}}\partial_{\bar{t}}\tilde{{\phi}}-
\tilde{{\mathcal L}}(\bar{t},{\bf y})
=\f{1}{2}
\left(\tilde{{\xi}}^2
+J|\nabla_{\bf x}\phi|^2\right)
+\tilde{{\xi}}(\partial_{\bar{t}}\tilde{{\phi}}-\sqrt{J}\partial_t\phi).
\end{eqnarray}
In the coordinates $(t,{\bf x})$, the Hamiltonian density  is given
by
\begin{eqnarray}
{{\mathcal H}}(t,{\bf x}) \equiv\tilde{{\mathcal H}} ({\mathcal
R}^{-1}(t,{\bf x}))\f{d^3{\bf y}}{d^3{\bf x}}
=\f{1}{J}\tilde{{\mathcal H}}({\mathcal R}^{-1}(t,{\bf x})).
\end{eqnarray}
Now, from expressions (\ref{(2)}) and (\ref{(6)}) we have that
$\tilde{{\xi}}(\bar{t},{\bf y})=\sqrt{J} {\xi}({\mathcal
R}(\bar{t},{\bf y})),$ and a straightforward calculation yields
\begin{eqnarray}
&&\hspace{-1cm}{{\mathcal H}}(t,{\bf x})= {{\mathcal E}}(t,{\bf x})
+\xi(t,{\bf x})<\partial_s{\bf R}({\mathcal R}^{-1}(t,{\bf x}))
,{\bf\nabla}_{\bf x}{{\phi}}(t,{\bf x})>+\f{1}{2} \left.\xi(t,{\bf
x}) \phi(t,{\bf x})
\partial_s(\ln J)\right|_{{\mathcal R}^{-1}(t,{\bf x})}.
\end{eqnarray}

\subsection{A simple and explicit example}

As a simple example, in the case of a single mirror following a prescribed
trajectory $(t,\epsilon g(t))$ in a $1+1$ space-time, we can take
$R(\bar{t},y)=y+\epsilon g(\bar{t})$, and thus we explicitly get
\begin{eqnarray}
{{\mathcal H}}(t, x)=
{{\mathcal E}}(t,x)
+\epsilon \dot{g}(t)\xi(t,x)\partial_x{{\phi}}(t, x).
\end{eqnarray}

\section{Single, partially transmitting mirror}

In this section we consider a single mirror in $1+1$ space-time
following a prescribed trajectory $(t,\epsilon g(t))$. When the
mirror is at rest, scattering of the field is described with the
$S-$matrix
\begin{eqnarray}S(\w)=\left(\begin{array}{cc}
{s}(\w)&{r}(\w)e^{-2i\w L}\\
{r}(\w)e^{2i\w L}&{s}(\w)\end{array}\right),\end{eqnarray} where
$x=L$ is the position of the mirror. The matrix $S$ is supposed to be
real in the temporal domain, as well as causal, unitary, and
transparent to high frequencies \cite{jr91}. More specifically,
these conditions appear naturally as a consequence of the following
considerations.
\begin{enumerate}
\item Since the field is neutral, it should be:
\begin{eqnarray} S(-\w)=S^*(\w).
\end{eqnarray}
In fact, the quantum field, in the Schr\"odinger picture,
 can be decomposed as
$$\hat{\phi}(x)=\sum_{j=R,L}\int_{\R}d\w\hat{a}_{\w,j}\tilde{g}_{\w,j}(x),
$$
where
\begin{eqnarray}&&
\tilde{g}_{\w,R}(x)=\f{1}{\sqrt{4\pi\w}}\left\{{s}(\w) e^{-i\w x}
\theta(L-x)+\left(e^{-i\w x}+{r}(\w)e^{-2i\w L}e^{i \w x}\right)
\theta(x-L)\right\},\label{modes0}\\&&\label{modes}
\tilde{g}_{\w,L}(x)=\f{1}{\sqrt{4\pi\w}}\left\{ \left(e^{i\w
x}+{r}(\w)e^{2i\w L}e^{-i \w x}\right)\theta(L-x) + {s}(\w)e^{i\w x}
\theta(L-x)\right\},
\end{eqnarray}
are  the right and the left incident modes, respectively \cite{bc95}. As is the usual
procedure in Quantum Field Theory, when $\w<0$ one performs the
change $\hat{a}_{\w,j}\rightarrow \hat{a}^{\dagger}_{-\w,j}$, and
thus, the field behaves as
$$\hat{\phi}(x)=\sum_{j=R,L}\int_{0}^{\infty}d\w
\left(\hat{a}_{\w,j}\tilde{g}_{\w,j}(x)+\hat{a}^{\dagger}_{\w,j}
\tilde{g}_{-\w,j}(x)\right).
$$
Now, since the field is neutral, it follows that
$\tilde{g}_{-\w,j}(x)=\tilde{g}_{\w,j}^*(x)$, and finally, we
conclude that
\begin{eqnarray*} S(-\w)=S^*(\w).
\end{eqnarray*} This proves the statement.
\item As a consequence of the commutation rule
$[{\hat{\phi}}(t,x),\hat{\phi}(t,y)]=0$, it follows that
\begin{eqnarray} S(\w)S^{\dagger}(\w)=Id. \label{(15)}
\end{eqnarray}
This is straightforward and needs no further comment.
\item
And, as a consequence of the commutation rule
$[{\hat{\xi}}(t,x),\hat{\phi}(t,y)]=-i\delta(x-y)$, we obtain the
following causality condition
$$\int_{\R}d\w \, r(w)e^{-i\w t}=\int_{\R}d\w \, s(w)e^{-i\w t}=0,
\quad \forall t<0,$$ in a distribution sense, i.e.,
$$\lim_{\gamma\rightarrow 0}\int_{\R}d\w \, r(w)\rho_{\gamma}(\w)e^{-i\w t}=
\lim_{\gamma\rightarrow 0}\int_{\R}d\w \, s(w)
\rho_{\gamma}(\w)e^{-i\w t}=0, \quad \forall t<0,$$ where
$\rho_{\gamma}$ is a frequency cut-off.
This condition is satisfied when
\begin{eqnarray}
\label{cau}
 S(\w) \mbox{ is analytic for Im}(\w)>0, \quad \mbox{ and }
s \mbox{ and } r \mbox{ are meromorphic functions.}
\end{eqnarray}
\item
A physical mirror is always transparent to high-enough incident
frequencies, thus it must necessarily hold
\begin{eqnarray} S(\w)\rightarrow Id, \quad \mbox{ when }
\quad |\w|\rightarrow \infty. \label{(17)}
\end{eqnarray}
\end{enumerate}

\subsection{Quantum theory based on the Hamiltonian approach}

In order to obtain the quantum theory, we will work in the coordinates
defined in the example above. In those coordinates the mirror is
situated at the point $y=0$ and the right and left incident modes
are given by Eqs.~(\ref{modes0}) and (\ref{modes}), with $L=0$.
Then, in the coordinates $(t,x)$, the instantaneous set of right and
the left incident eigenfunctions, which generalize the set used in the
case of a perfectly reflecting mirror, is
\begin{eqnarray}
g_{\w,j}(t,x;\epsilon)\equiv \tilde{g}_{\w,j}(x-\epsilon g(t))
\qquad j=R,L.
\end{eqnarray}
Note that, in general, we do not know explicitly the part of the Hamiltonian
that describes the interaction between the field and the mirror. As
a consequence, in order to obtain the quantum theory, the energy of
the field $E(t)=\int_{\R}dx\, {\mathcal E}(t,x)$, which in the presence
of a single mirror does not depend on $\epsilon$, should be viewed as
part to the free Hamiltonian of the system.

As is usual, working in the interaction picture, the field is expanded as
follows:
\begin{eqnarray}&&\label{field}
\hat{\phi}_I(t,x;\epsilon)=\sum_{j=R,L}\int_0^{\infty}d\w\hat{a}_{w,
j}e^{-i\w t} g_{\w,j}(t,x;\epsilon)+hc,\nonumber\\&&
\hat{\xi}_I(t,x;\epsilon)=-i\sum_{j=R,L}\int_0^{\infty}d\w
\w\hat{a}_{w, j} e^{-i\w t} g_{\w,j}(t,x;\epsilon)+hc,
\end{eqnarray}
where $hc$ means, in each case, the Hermitian conjugate of the preceding
expression. The quantum equation, in this picture, is given by
\begin{eqnarray}&&\hspace{-1cm}\label{sch}
i\partial_t|\Phi\rangle=
\f{\epsilon \dot{g}(t)}{2}\left[\int_{\R}dx\hat{\xi}_I(t,x;\epsilon)
\partial_x\hat{\phi}_I(t, x;\epsilon)+
hc\right]
|\Phi\rangle\nonumber\\&&=
\f{\epsilon \dot{g}(t)}{2}\left[\int_{\R}dx\hat{\xi}_I(t,x;0)
\partial_x\hat{\phi}_I(t, x;0)+
hc\right] |\Phi\rangle+{\mathcal O}(\epsilon^2).
\end{eqnarray}
Let now ${\mathcal T}^t$ be the quantum evolution operator of the
Schr\"odinger equation $(\ref{sch})$, and let  $|0\rangle$ be the
initial quantum state. Then, the average number of produced particles
 at time $t$ is
\begin{eqnarray}
{\mathcal N}(t)\equiv\sum_{j=R,L}\int_0^{\infty}d\w \, \langle
0|\left({\mathcal T}^{t}\right)^{\dagger}
\hat{a}^{\dagger}_{\w,j}\hat{a}_{\w,j} {\mathcal T}^{t}\, |0\rangle,
\end{eqnarray}
and the dynamical energy at time $t$, that is the energy of the
created particles at time $t$, is obtained as
\begin{eqnarray}
\langle\hat{E}(t)\rangle\equiv\sum_{j=R,L}\int_0^{\infty}d\w\, \w\,
\langle 0|\left({\mathcal T}^{t}\right)^{\dagger}
\hat{a}^{\dagger}_{\w,j}\hat{a}_{\w,j} {\mathcal T}^{t}\, |0\rangle.
\end{eqnarray}
We should note that, during the movement of the mirror, the particles have been
called sometimes quasi-particles, owing in part to the difficulties encountered
in the past when trying to give them a physical sense (see \cite{gmm94}).

A simple but rather cumbersome calculation yields the following results
\begin{eqnarray}\label{num}
{\mathcal N}(t)=\f{\epsilon^2}
{2\pi^2}
\int_0^{\infty}\int_0^{\infty}\f{d\w d\w'\w \w'}
{ (\w+\w')^2}\left|
\widehat{\dot{g}\theta_t} (\w+\w')\right|^2
(|{r}(\w)+{r}^*(\w')|^2+|{s}(\w)-{s}^*(\w')|^2)
+{\mathcal O}(\epsilon^4),
\end{eqnarray}
\begin{eqnarray}\label{ene}
\langle\hat{E}(t)\rangle=\f{\epsilon^2}
{4\pi^2}
\int_0^{\infty}\int_0^{\infty}\f{d\w d\w'\w \w'}
{ \w+\w'}\left|
\widehat{\dot{g}\theta_t}(\w+\w')\right|^2
(|{r}(\w)+{r}^*(\w')|^2+|{s}(\w)-{s}^*(\w')|^2)
+{\mathcal O}(\epsilon^4),
\end{eqnarray}
where $\theta_t$ is the Heaviside step function at point $t$, e.g.,
$\theta_t(\tau)=\theta(t-\tau)$,
 and  $\hat{f} (\w) \equiv
\int_{\R}d\tau f(\tau) e^{i\w \tau} $ is the Fourier transformed of
the function  $f$. These two quantities are in general convergent.
However for the Davis-Fulling model \cite{fd76} ---that is, in the
case of a single perfectly reflecting mirror--- such quantities are
divergent when the mirror moves or when the displacement exhibits some type
of discontinuity \cite{mo70,sps98}.

In this situation, in order to obtain a finite energy, several
authors have used different regularization techniques
\cite{fv82}-\cite{e93}. For example
using a frequency cut-off  $e^{-\w\gamma}$ with $0<\gamma\ll 1$, the
regularized energy is (see Ref.~\cite{haro1})
\begin{eqnarray}
\langle\hat{E}(t;\gamma)\rangle= \f{\epsilon^2}{6\pi}\left[
\f{\dot{g}^2(t)}{\pi\gamma}-\ddot{g}(t)\dot{g}(t)+
\int_0^t\ddot{g}^2(\tau)d\tau\right] + {\cal O} (\epsilon^4) .\end{eqnarray}
 Thus, imposing the kinetic energy of the moving boundary to be
\begin{eqnarray}
\f{1}{2}
\left(M_{exp}-\f{1}{3\pi^2\gamma}\right)\epsilon^2\dot{g}^2(t),
\end{eqnarray}
where $M_{exp}$ is the experimental mass of the mirror, those
authors conclude that the renormalized dynamical energy, namely
$\hat{E}_R(t)$, is (see Refs.~\cite{fd76}, \cite{fv82}-\cite{e93})
\begin{eqnarray}
\langle\hat{E}_R(t)\rangle \equiv
\f{\epsilon^2}{6\pi}\left[-\ddot{g}(t)\dot{g}(t)+
\int_0^t\ddot{g}^2(\tau)d\tau\right]+ {\cal O} (\epsilon^4).
\end{eqnarray} However, when
$t\leq \delta$, with $0<\delta\ll 1$ such renormalized energy is
negative. This shows that, when the mirror moves, the renormalized
energy cannot be considered as the  energy of the produced particles
at time $t$ (see also the paragraph immediately after Eq.~(4.5) in
Ref.~\cite{fd76}).

From our viewpoint, such meaningless result is just due to the fact
that a perfect reflecting mirror is used in the derivation, what is
not physically feasible at any price. Physical mirrors will {\it always}
obey a transparency condition of the kind here proposed (\ref{(17)}), and
then it comes out for free that the
average number of produced particles and the dynamical energy turn
out to be {\it well defined} and are both {\it positive} quantities.

We have also calculated the radiation-reaction force. For a single
mirror this force is the difference between the energy density of
the evolved vacuum state on both sides of the mirror, namely,
\begin{eqnarray}
\langle\hat{F}_{Ha}(t)\rangle\equiv\lim_{\delta\rightarrow 0}\left(
\langle 0|\left({\mathcal T}^{t}\right)^{\dagger}
\hat{\mathcal E}(t,\epsilon g(t)-|\delta|)
{\mathcal T}^{t}|0\rangle
-\langle 0|\left({\mathcal T}^{t}\right)^{\dagger}
\hat{\mathcal E}(t,\epsilon g(t)+|\delta|)
{\mathcal T}^{t}|0\rangle\right),
\end{eqnarray}
where the subindex {\it Ha} means that the radiation-reaction force
has been calculated in the Hamiltonian approach. We obtain
\begin{eqnarray}&&\hspace{-1cm}\label{fha}
\langle\hat{F}_{Ha}(t)\rangle =-\f{\epsilon} {2\pi^2}
\int_0^{\infty}\int_0^{\infty}\f{d\w d\w'\w \w'} {\w+\w'}\
\Re\left[e^{-i(\w+\w')t}
\widehat{\dot{g}\theta_t}(\w+\w')\right]\nonumber
\\&&\hspace{2cm}\times
(|{r}(\w)+{r}^*(\w')|^2+|{s}(\w)-{s}^*(\w')|^2)
+{\mathcal O}(\epsilon^2).
\end{eqnarray}
An important remark is here in order. Note that, as a consequence of
the energy conservation law, the dynamical energy at time $t$ is
equal to minus the work performed by the
radiation-reaction force up to time $t$ (see Refs.~\cite{fv82} and \cite{be94}).
This law is naturally satisfied if we use the Hamiltonian approach.
It is then clear that (\ref{ene}) and (\ref{fha}) are related through the
formula
\begin{eqnarray}
\langle\hat{E}(t)\rangle=-\epsilon\int_0^t\langle\hat{F}_{Ha}(\tau)
\rangle\dot{g}(\tau)d\tau.\end{eqnarray}

\subsection{Quantum theory in the Heisenberg picture}

Following the method of \cite{op01}, we have calculated the ``in''
modes when the mirror describes the prescribed trajectory
$(t,\epsilon g(t))$. Using light-like coordinates, $u\equiv t+x$
and $v\equiv t-x$, the ''in'' modes can be written as
\begin{eqnarray}&&\hspace{-1cm}
\phi^{in}_{\w,R}(u,v;0)=\f{1}{\sqrt{4\pi\w}}\left\{\left[{s}(\w)e^{-i\w
v} -A_{\w}(v;0) \right] \theta(\epsilon
g(t)-x)\nonumber\right.\\&&\left. +\left[e^{-i\w v}+{r}(\w)e^{-i\w
u} -B_{\w}(u;0) \right] \theta(x-\epsilon g(t)) \right\}+{\mathcal
O}(\epsilon^2),
\end{eqnarray}
\begin{eqnarray}&&\hspace{-1cm}
\phi^{in}_{\w,L}(u,v;0)=\f{1}{\sqrt{4\pi\w}}\left\{\left[ e^{-i\w
u}+{r}(\w)e^{-i\w v} +B_{\w}(v;0) \right] \theta(\epsilon
g(t)-x)\nonumber\right.\\&&\left.\hspace{0.5cm}
+\left[{s}(\w)e^{-i\w u} +A_{\w}(u;0) \right] \theta(x-\epsilon
g(t)) \right\}+{\mathcal O}(\epsilon^2),
\end{eqnarray}
where
$$A_{\w}(y;\gamma)=
\f{i\epsilon\w}{2\pi}\int_{\R}d\w'e^{i\w'y}\hat{g}(-\w-\w')
[{s}^*(\w')-{s}(\w)]e^{-\gamma|\w'|},
$$
$$B_{\w}(y;\gamma)=
\f{i\epsilon\w}{2\pi}\int_{\R}d\w'e^{i\w' y}\hat{g}(-\w-\w')
[{r}^*(\w')+{r}(\w)]e^{-\gamma|\w'|}.
$$
The average number of produced particles after the mirror returns to
rest is \cite{bd82}
\begin{eqnarray}
{\mathcal N}(t\geq T)=\sum_{i,j=R,L}
\int_0^{\infty}
\int_0^{\infty}d\w d\w'\left|(\phi^{out }_{\w,i},\phi^{in *}_{\w',j})
\right|^2,
\end{eqnarray}
where $(F,G)\equiv i\int_{\R}
dx(F^*\partial_tG-G\partial_tF^* )$.
To obtain an explicit result, we calculate the
  Bogoliubov coefficients
$\left\{\phi^{out }_{\w,i},\phi^{in *}_{\w',j}\right\}$
 in the null future infinity ${\mathcal I}^+$, because the ``out''-going
modes are a very simple expression in ${\mathcal I}^+$. The final
result is (see Ref.~\cite{jr9236}):
\begin{eqnarray}&&\hspace{-1cm}\label{tnum}
{\mathcal N}(t\geq T)=\f{\epsilon^2}
{2\pi^2}
\int_0^{\infty}\int_0^{\infty}d\w d\w'\w\w'\left|
\hat{g}(\w+\w')\right|^2
(|{r}(\w)+{r}^*(\w')|^2+|{s}(\w)-{s}^*(\w')|^2)
+{\mathcal O}(\epsilon^4).
\end{eqnarray}
From this expression it is not difficult to calculate the number of
particles at time $t$. We only need to consider the function
\begin{eqnarray*}
\tilde{g}_t(s)\equiv\left\{\begin{array}{ccc}
g(s),& \mbox{when}& s\leq t,\\
g(t),& \mbox{when}& s\geq t,
\end{array}\right.\end{eqnarray*}
because  $\hat{\tilde{g}}_t(\w+\w')=\f{1}{\w+\w'}
\widehat{\dot{g}\theta_t}(\w+\w')$. Then, inserting this expression
into $(\ref{tnum})$ we obtain formula $(\ref{num})$.

The radiation-reaction force calculated in the Heisenberg picture,
namely $\langle\hat{F}_H(t)\rangle$, is the difference between the
energy density of the ``in'' vacuum state on the left and on the
right side of the mirror. A simple calculation shows that the energy
density of the ``in'' vacuum \cite{bd82}
$$\langle\hat{\mathcal E}(t,x)\rangle=\sum_{j=R,L}
\int_0^{\infty}d\w\left(\partial_u\phi^{in}_{\w,j}
(u,v;0)\partial_u\phi^{in *}_{\w,j}(u,v;0)+
\partial_v\phi^{in}_{\w,j}(u,v;0)
\partial_v\phi^{in *}_{\w,j}(u,v;0)\right),$$
on both sides of the mirror, is
\begin{eqnarray}&&\hspace{-2cm}
\langle\hat{\mathcal E}(t,x)\rangle
=\int_0^{\infty}d\w \w
\pm \f{i\epsilon}{8\pi^2}
\int_{\R^2}d\w d\w' \w \w' \hat{g}(\w+\w')
\chi(\w) \nonumber\\&&\hspace{1cm} \times
(1+r(\w)r(\w')-s(\w)s(\w'))e^{-i(\w+\w')v}
\theta(\pm(\epsilon g(t)-x))+{\mathcal O}(\epsilon^2),
\end{eqnarray}
where $\chi(\w)\equiv\theta(\w)-\theta(-\w)$ is the sign function.
Note that the term of order $\epsilon$ is ill-defined, because the
function $\w \w' \hat{g}(\w+\w')(1+r(\w)r(\w')-s(\w)s(\w'))$ is not
integrable in $\R^2$ and, to obtain a well-defined quantity,
appropriate regularization is needed.

If we define the regularized energy by
\begin{eqnarray}
\langle\hat{\mathcal E}(t,x;\gamma)\rangle\equiv\sum_{j=R,L}
\int_0^{\infty}d\w
e^{-\gamma\w}
\left(\partial_u\phi^{in}_{\w,j}
(u,v;\gamma)\partial_u\phi^{in *}_{\w,j}(u,v;\gamma)+
\partial_v\phi^{in}_{\w,j}(u,v;\gamma)
\partial_v\phi^{in *}_{\w,j}(u,v;\gamma)\right),
\end{eqnarray}
then the regularized motion force, in the Heisenberg picture results
\begin{eqnarray}&&
\langle\hat{F}_H(t;\gamma)\rangle=\f{i\epsilon}{8\pi^2}
\int_{\R^2}d\w d\w' \w \w' \hat{g}(\w+\w') (\chi(\w)+\chi(\w'))
\nonumber\\&&\hspace{2cm}\times(1+r(\w)r(\w')-s(\w)s(\w'))
e^{-\gamma(|\w|+|\w'|)} e^{-i(\w+\w')t}+{\mathcal O}(\epsilon^2).
\end{eqnarray}
This integral is convergent and already cut-off {\it independent}. Thus, a
natural definition of the
 renormalized radiation-reaction force is
\begin{eqnarray}&&\label{fhei}
\langle\hat{F}_{H,ren}(t)\rangle=\f{i\epsilon}{8\pi^2}
\int_{\R^2}d\w d\w' \w \w' \hat{g}(\w+\w') (\chi(\w)+\chi(\w'))
\nonumber\\&&\hspace{2cm}\times(1+r(\w)r(\w')-s(\w)s(\w'))
e^{-i(\w+\w')t}+{\mathcal O}(\epsilon^2).
\end{eqnarray}

Two more remarks are in order. In general, expression
$(\ref{fhei})$ disagrees with the radiation-reaction force
$(\ref{fha})$ obtained using the Hamiltonian approach. A detailed
discussion of this point will be given in the next section.

It is also important to stress that {\it other} definitions of the
regularized energy density give rise to {\it different}
motion forces. It is, of course, also possible to obtain the
radiation-reaction force (\ref{fha}) within a specific, tailored
regularization. To prove such statement let us consider, for a moment,
the usual case in the literature of a
perfectly reflecting mirror. In the Heisenberg picture the
annihilation operators on the left side of the mirror are
\begin{eqnarray*}&&
\hat{a}_{\w,L}(t)=\hat{a}_{\w,L}e^{-i\w t}
+ e^{-i\w t}\left[\epsilon\int_0^{\infty}d\w'
\widehat{\dot{g}\theta_t}(\w+\w')\hat{a}_{\w',L}^{\dagger}
\f{\sqrt{\w\w'}}{\pi}\f{1}{\w+\w'}\right.\\&&\left.\hspace{2cm}
-\epsilon\int_0^{\infty}d\w'\widehat{\dot{g}\theta_t}(\w-\w')
\hat{a}_{\w',L}
\f{\sqrt{\w\w'}}{\pi}{\mathcal P}\left(\f{1}{\w-\w'}\right)\right]
+\mathcal{O}(\epsilon^2),
\end{eqnarray*}
where ${\mathcal P}$ denotes Cauchy's principal value.
Then, on the left side of the mirror, the field ---in the Heisenberg
picture--- can be written as follows
\begin{eqnarray*}
\hat{\phi}_H(t,x;\epsilon)=\int_0^{\infty}d\w\left(
\hat{a}_{\w,L}(t)g_{\w,L}(t,x;\epsilon)+
\hat{a}_{\w,L}^{\dagger}(t)g_{\w,L}^*(t,x;\epsilon)\right)
\end{eqnarray*}
and, after some algebra, we get
\begin{eqnarray*}
\hat{\phi}_H(t,x;\epsilon)=\int_0^{\infty}d\w\left(
\hat{a}_{\w,L}\varphi_{\w,L}(x,t;0)+
\hat{a}_{\w,L}^{\dagger}\varphi_{\w,L}^*(x,t;0)\right),
\end{eqnarray*}
where
\begin{eqnarray*}&&
\varphi_{\w,L}(t,x;\gamma)= \f{ie^{-i\w t}}{\sqrt{\pi \w}}
\left[\sin(\w x)-\epsilon g(t)\cos(\w x)\right]\\&&\hspace{2cm}
-\f{i\epsilon\sqrt{\w}}{\pi\sqrt{\pi}} \int_{\R}d\w'd\tau e^{i\w' t}
\widehat{\dot{g}\theta_t}(-\w-\w') {\mathcal
P}\left(\f{1}{\w+\w'}\right)\sin(\w' x) e^{-\gamma|\w'|}
+\mathcal{O}(\epsilon^2).
\end{eqnarray*}
Using the formula
\begin{eqnarray*}
\int_{\R}d\w'{\mathcal P}\left(\f{1}{\w+\w'}\right)e^{i\w'(u-\tau)}=
\pi i e^{-i\w(u-\tau)}\chi(u-\tau),
\end{eqnarray*}
we easily find that
$$\phi^{in}_{\w,L}(t,x;0)=
\f{i}{\sqrt{\pi\w}}\sin(\w x)e^{-i\w
t}-2i\epsilon\sqrt{\f{\w}{4\pi}} e^{-i\w(t-x)}g(t-x)=
\varphi_{\w,L}(t,x;0),$$ and making the changes $\epsilon\rightarrow
-\epsilon$ and $x\rightarrow -x$, we obtain the expression for the
right modes. That is, we have obtained an equivalent expression as for
the ``in'' modes.

 Then, defining the regularized energy from this new expression of the
``in'' modes
\begin{eqnarray*}
\langle\hat{\mathcal E}(t,x;\gamma)\rangle\equiv\f{1}{2}\sum_{j=R,L}
\int_0^{\infty}d\w
e^{-\gamma\w}
\left(\partial_t\varphi_{\w,j}
(t,x;\gamma)\partial_t\varphi^{*}_{\w,j}(t,x;\gamma)+
\partial_x\varphi_{\w,j}(t,x;\gamma)
\partial_x\varphi^{*}_{\w,j}(t,x;\gamma)\right),
\end{eqnarray*}
we obtain the following regularized motion force
\begin{eqnarray*}&&\hspace{-1cm} \langle\hat{F}_{H}(t;\gamma)\rangle =-\f{2\epsilon}
{\pi^2}
\int_0^{\infty}\int_0^{\infty}\f{d\w d\w'\w \w'}
{\w+\w'}\
\Re\left[e^{-i(\w+\w')t}
\widehat{\dot{g}\theta_t}(\w+\w')\right]e^{-\gamma(|\w|+|\w'|)}
+\mathcal{O}(\epsilon^2),
\end{eqnarray*}
which agrees with $(\ref{fha})$, for the case of a perfectly
reflecting mirror.

The fact that different (sometimes {\it a priori} quite reasonable)
regularization procedures may lead to very different finite results is
well known \cite{eea}, as also the ensuing consequence that there is
generically no physics associated with an arbitrary regularization
prescription, which calls for a subsequent renormalization procedure to
establish contact with the physical world. This is more so when one deals with
plain mathematical and physically unrealistic boundary conditions, as we have
learnt from a number of situations involving the ordinary Casimir effect too
\cite{jaffeea}. However, this essential point seems to have been put aside,
at least to some extent, when dealing with the problem at hand, maybe due to
the intrinsic mathematical difficulty of this issue here. This
was the motivation for the last explicit calculation above, that we consider a
sufficiently clarifying exercise which exhibits what is going on here.
In the following we will proceed with a strict comparison of our
results with those of other authors that have previously appeared in the literature
on the subject.

\subsection{Comparison with known results}

\noindent {\sl (i) The method of Jaekel and Reynaud.} \medskip

To study the radiation-reaction force, these authors \cite{jr92}
consider  the following effective Hamiltonian
\begin{eqnarray}
\hat{H}_{J,R}\equiv -\epsilon g(t)\hat{F}(t),
\end{eqnarray}
where $(t,\epsilon g(t))$ is the trajectory of the mirror and
$\hat{F}(t)\equiv \lim_{\delta\rightarrow 0} \left(\hat{\mathcal
E}(t,-|\delta|)- \hat{\mathcal E}(t,+|\delta|)\right)$ is the force
operator at the point $x=0$. A simple calculation yields
\begin{eqnarray}&&\label{fjr}
\langle\hat{F}_{J,R}(t)\rangle=-\f{\epsilon} {2\pi^2}
\int_0^{\infty}\int_0^{\infty}d\w d\w'\w\w'\ \Im\left[e^{-i(\w+\w')t}
\widehat{{g}\theta_t}(\w+\w')\right]\nonumber\\&&\hspace*{2cm}\times
(|{r}(\w)+{r}^*(\w')|^2+|{s}(\w)-{s}^*(\w')|^2)+{\mathcal
O}(\epsilon^2).
\end{eqnarray}
Integrating by parts, we obtain
\begin{eqnarray*}&&
\langle\hat{F}_{J,R}(t)\rangle=\langle\hat{F}_{Ha}(t)\rangle
+\f{\epsilon g(t)}{2\pi^2}\int_0^{\infty}\int_0^{\infty}
\f{d\w d\w'\w\w'}{\w+\w'}
(|{r}(\w)+{r}^*(\w')|^2+|{s}(\w)-{s}^*(\w')|^2)+{\mathcal O}(\epsilon^2),
\end{eqnarray*}
what shows that expression $(\ref{fjr})$ is
divergent while the mirror moves. To obtain a regularized quantity
we write $(\ref{fjr})$ as follows
\begin{eqnarray*}&&\hspace{-1cm}
\langle\hat{F}_{J,R}(t)\rangle\equiv\f{i\epsilon}
{8\pi^2}\int_{-\infty}^{t}d\tau g(\tau) \int_{\R^2}d\w
d\w'\w\w'(\chi(\w)+\chi(\w')) \left[(1+r(\w)r(\w')-s(\w)s(\w'))
\right.\nonumber\\&&\hspace*{1cm}\left.+(1+r^*(\w)r^*(\w')-s^*(\w)s^*(\w'))\right]
e^{-i(\w+\w')(t-\tau)}+{\mathcal O}(\epsilon^2),
\end{eqnarray*}
and we define the regularized motion force by
\begin{eqnarray}&&\hspace{-1cm}
\langle\hat{F}_{J,R}(t;\gamma)\rangle\equiv\f{i\epsilon}
{8\pi^2}\int_{-\infty}^{t}d\tau g(\tau) \int_{\R^2}d\w
d\w'\w\w'(\chi(\w)+\chi(\w'))
\left[(1+r(\w)r(\w')-s(\w)s(\w'))\rho_{\gamma}(\w,\w')
\right.\nonumber\\&&\hspace*{1cm}\left.+(1+r^*(\w)r^*(\w')-s^*(\w)s^*(\w'))
\rho_{\gamma}^*(\w,\w')\right] e^{-i(\w+\w')(t-\tau)}+{\mathcal
O}(\epsilon^2),
\end{eqnarray}
where  the cut-off  $\rho_{\gamma}(\w,\w')$ is a meromorphic
function, analytic for Im$(\w)>0$ and Im$(\w')>0$.

Now, applying the causality of the $S$-matrix (Eq.~(\ref{cau})), and
making $\gamma\rightarrow 0$, an easy calculation leads us to the
expression (\ref{fhei}). For this reason, defining the renormalized
radiation-reaction force through the formula $(\ref{fhei})$, i.e.,
\begin{eqnarray}
\langle\hat{F}_{J,R,ren}(t)\rangle\equiv
\langle\hat{F}_{H,ren}(t)\rangle,
\end{eqnarray}
one does conclude that the method of Jaekel and Reynaud is
equivalent to the quantum theory in the Heisenberg picture.

Note also that $$\epsilon\int_{\R}dt\, \langle\hat{F}_{Ha}(t)\rangle
\dot{g}(t)= \epsilon\int_{\R}dt\, \langle\hat{F}_{J,R,ren}(t)\rangle
\dot{g}(t).$$ This identity proves that the dissipative parts of
$\langle\hat{F}_{Ha}(t)\rangle$ and
$\langle\hat{F}_{J,R,ren}(t)\rangle$ always agree.

On the other hand, in several situations, the reactive part
disagrees actually. For example, if
${r}(w)=-\f{i\alpha}{\w+i\alpha}$, and ${s}(w)=\f{\w}{\w+i\alpha}$,
with $\alpha>0$, there holds the relation
\begin{eqnarray}
\langle\hat{F}_{Ha}(t)\rangle=-\f{\alpha\epsilon}{2\pi}\ddot{g}(t)
+
\langle\hat{F}_{J,R,ren}(t)\rangle,
\end{eqnarray}
where
\begin{eqnarray}
\langle\hat{F}_{J,R,ren}(t)\rangle=\f{\alpha\epsilon}{\pi}
\int_1^{\infty}dz\int_{-\infty}^td\tau\left(\f{1}{z^2}-\f{1}{z^3}\right)
e^{-\alpha z(t-\tau)}\dddot{g}(\tau).
\end{eqnarray}
That is, both motion forces differ in a reactive term. Note also
that, during the oscillation of the mirror, the work done by the motion
force $\langle\hat{F}_{J,R,ren}(t)\rangle$ is {\it not} a negative
quantity. Consequently, from the previous remark it follows that the
dynamical energy is not  positive, and therefore a seemingly meaningless
result is obtained since, in our opinion, the dynamical energy is
to be interpreted as the energy carried out by the produced particle.
To avoid such difficulty the reactive term
$-\f{\alpha\epsilon}{2\pi}\ddot{g}(t)$
should not be arbitrarily suppressed but, on the contrary, has to be
duely taken into account. This saves the day and endows the whole picture
with physical sense, as explained in the previous section.\medskip

\noindent{\sl (ii) The method of Barton and Calogeracos.}\medskip

In Ref.~\cite{bc95} (see also \cite{sb95} and \cite{dmn1}), these
authors study the particular case ${r}(w)=-\f{i\alpha}{\w+i\alpha}$,
and ${s}(w)=\f{\w}{\w+i\alpha}$ with $\alpha>0$. In such situation,
the interaction between the field and the mirror can be described by
the Lagrangian density
\begin{eqnarray}\label{lbc}
\f{1}{2}\left[(\partial_t\phi)^2-(\partial_x\phi)^2\right]
-\alpha\phi^2 \delta(x-\epsilon g(t)).
\end{eqnarray}
Following the method discussed in Sect.~II, we have obtained the
quantum Hamiltonian
\begin{eqnarray}
\hat{H}(t)=\int_{\R}dx\ \hat{{\mathcal E}}(t,x)
+\alpha\hat{\phi}^2(t,\epsilon g(t)) +\frac{\epsilon
\dot{g}(t)}{2}\left[\int_{\R}dx\
\hat{\xi}(t,x)\partial_x\hat{\phi}(t, x)+hc\right].
\end{eqnarray}

Now, inserting (\ref{field}) into the integral
$\int_{\R}dx\, \hat{{\mathcal E}}(t,x) +\alpha\hat{\phi}^2(t,\epsilon
g(t))$, we get
\begin{eqnarray}
\int_{\R}dx\ \hat{{\mathcal E}}(t,x) +\alpha\hat{\phi}^2(t,\epsilon
g(t))= \sum_{j=L,R}\int_0^{\infty}d\w \ \w
\left(\hat{a}^{\dagger}_{\w,j} \hat{a}_{\w,j}+\f{1}{2}\right).
\end{eqnarray}
We thus conclude that the quantum equation in the interaction
picture is given by expression (\ref{sch}). And, consequently,
for these reflection and transmission coefficients, the authors would
obtain the same formulae (\ref{num}), (\ref{ene}) and (\ref{fha}).

It should be noted, however, that two relevant differences exist
between their results and the ones we have derived here previously.
\begin{enumerate}
\item In order to obtain the quantum equation, B-C make a
unitary transformation which does not seem to be easily generalizable to
the case of two moving mirrors. In our case this can be done without the least
problem, as we shall show below.
\item In the above mentioned paper, Ref.~\cite{bc95}, the authors use
the same technique for mass renormalization that had been employed by
Barton and Eberlein in Refs.~\cite{be94} and
\cite{e93} for the case of a completely reflecting mirror, in order to
eliminate the reactive part of the motion force. However, within
such renormalization, the energy of the field is not a positive
quantity for all time $t$ and, consequently, the concept of particle
is again ill-defined during the oscillation of the mirror.
\end{enumerate}

\section{Two partially transmitting mirrors}
In this section we consider the situation where we have two moving
mirrors which follow prescribed trajectories
$(t,L_j(t;\epsilon))$, where $L_j(t;\epsilon)\equiv L_j+\epsilon
g_j(t)$, with $j=1,2$, and we assume that
$L_1(t;\epsilon)<L_2(t;\epsilon),$ $\forall t\in \R$. In this case
it is impossible, in practice, to work within the Heisenberg picture,
because it is indeed very complicated to obtain the ``in'' and
``out'' mode functions in the presence of the two moving mirrors.
Alternatively, with the purpose to calculate the dissipative part of
the motion force, the number of radiated particles, and their
energy, one can use the approach due to Jaekel and Reynaud and                                          based on the
effective Hamiltonian $\hat{H}_{J,R}\equiv -\sum_{j=1,2}\epsilon
g_j(t)\hat{F}_j(t)$, where $\hat{F}_j(t)\equiv
\lim_{\delta\rightarrow 0} \left(\hat{\mathcal E}(t,L_j-|\delta|)-
\hat{\mathcal E}(t,L_j+|\delta|)\right)$ is the force operator at
the point $x=L_j$ (see Refs.~\cite{jr92} and \cite{jr9236}). However
this method is not useful in order to calculate the reactive part of
the motion force or the dynamical energy while the mirrors are in
movement.

To obtain those last quantities we are led to use, once more, the
Hamiltonian approach. In this case, if we consider the change
$$R(\bar{t},y)= \f{1}{L_2-L_1}\left[
L_2(\bar{t};\epsilon)(y-L_1)+L_1(\bar{t};\epsilon)(L_2-y)\right],$$
the Hamiltonian density of the field is
\begin{eqnarray}
{\mathcal H}(t,x)={\mathcal E}(t,x)
+\sum_{j=1,2}\f{(-1)^j\dot{L}_j(t;\epsilon)\xi(t,x)}{
L_2(t;\epsilon)-L_1(t;\epsilon)}\left[
\partial_x\phi(t,x)(x-\bar{L}_{j}(t;\epsilon))
+\f{1}{2}\phi(t,x)\right],
\end{eqnarray}
where $\bar{L}_{\left(\substack{ 1\\2}\right)}(t;\epsilon)\equiv
{L}_{\left(\substack{ 2\\1}\right)}(t;\epsilon)$.
In the coordinates $(\bar{t},y)$ the set of right and left incident
modes can be obtained from Eqs.~(8) and (9) of Ref.~\cite{jr9236}.
We find
\begin{eqnarray}&&\hspace{-1cm}
\tilde{g}_{\w,R}(y)=\f{1}{\sqrt{4\pi\w}}
\left\{\f{{s}_1(\w){s}_2(\w)}{d(\w)}
e^{-iwy}
\theta(L_1-y)
\right.\nonumber\\&&\left.
+\left(\f{{s}_2(\w)}{d(\w)}e^{-iwy}+
\f{{r}_1(\w){s}_2(\w)}{d(\w)}
e^{iw(y-2L_1)}\right)
\theta(y-L_1)\theta(L_2-y)
\right.\nonumber\\&&\left.
+\left[e^{-iwy}+\left({r}_2(\w)e^{-2iwL_2}+
\f{{r}_1(\w){s}^2_2(\w)}{d(\w)}e^{-2iwL_1}\right)
e^{iwy}\right]\theta(y-L_2)
\right\},
\end{eqnarray}
\begin{eqnarray}&&\hspace{-1cm}
\tilde{g}_{\w,L}(y)=\f{1}{\sqrt{4\pi\w}}
\left\{\left[e^{iwy}+\left({r}_1(\w)e^{2iwL_1}+
\f{{r}_2(\w){s}^2_1(\w)}{d(\w)}e^{2iwL_2}\right)
e^{-iwy}\right]\theta(L_1-y)
\right.\nonumber\\&&\left.
+\left(\f{{s}_1(\w)}{d(\w)}e^{iwy}+
\f{{r}_2(\w){s}_1(\w)}{d(\w)}
e^{-iw(y-2L_2)}\right)
\theta(y-L_1)\theta(L_2-y)
\right.\nonumber\\&&\left.
+\f{{s}_1(\w){s}_2(\w)}{d(\w)}
e^{iwy}
\theta(y-L_2)
\right\},
\end{eqnarray}
where $d(\w)\equiv 1-r_1(\w)r_2(\w)e^{2i\w(L_2- L_1)}$. Then, the
instantaneous set of right and left incident eigenfunctions in the
coordinates $(t,x)$ is
\begin{eqnarray}
{g}_{\w,R}(t,x;\epsilon)=
\sqrt{\f{L_2-L_1}{L_2(t;\epsilon)-L_1(t;\epsilon)}}
\tilde{g}_{\w,R}(y(t,x)),\quad
{g}_{\w,L}(t,x;\epsilon)=
\sqrt{\f{L_2-L_1}{L_2(t;\epsilon)-L_1(t;\epsilon)}}
\tilde{g}_{\w,L}(y(t,x)).
\end{eqnarray}
The fields can by expanded as follows
\begin{eqnarray}&&
\hat{\phi}(t,x)= \sum_{j=R,L}\int_0^{\infty}d\w
\hat{a}_{\w,j}{g}_{\w,j}(t,x;\epsilon)+hc, \nonumber\\&&
\hat{\xi}(t,x)= -i\sum_{j=R,L}\int_0^{\infty}d\w
\w\hat{a}_{\w,j}{g}_{\w,j} (t,x;\epsilon)+hc.
\end{eqnarray}
In this case the energy of the fields depends on $\epsilon$. In fact,
we have
\begin{eqnarray}&&
\hat{E}(t)\equiv\int_{\R}dx \hat{{\mathcal
E}}(t,x)=\f{1}{2}\int_{\R}dy
\left[\left(\widehat{\tilde{\xi}}(y)\right)^2+
\left(\partial_y{\widehat{\tilde{\phi}}}(y)\right)^2\right]
\nonumber\\&&\hspace{1cm}-
\f{\epsilon(g_2(t)-g_1(t))}{L_2-L_1}\int_{\R}dy
\left(\partial_y{\widehat{\tilde{\phi}}}(y)\right)^2+ {\mathcal
O}(\epsilon^2),
\end{eqnarray}
where we have introduced the ``free'' fields (the fields when the
two mirrors are at rest)
\begin{eqnarray}&&
\widehat{\tilde{\phi}}(y)= \sum_{j=R,L}\int_0^{\infty}d\w
\hat{a}_{\w,j}\tilde{g}_{\w,j}(y)+hc, \nonumber\\&&
\widehat{\tilde{\xi}}(y)= -i\sum_{j=R,L}\int_0^{\infty}d\w
\w\hat{a}_{\w,j}\tilde{g}_{\w,j} (y)+hc.
\end{eqnarray}
In the same way, the corresponding quantum Hamiltonian is obtained as
\begin{eqnarray}&&
\hat{H}(t)\equiv\int_{\R}dx\hat{\mathcal H}(t,x)=
\f{1}{2}\int_{\R}dy \left[\left(\widehat{\tilde{\xi}}(y)\right)^2+
\left(\partial_y{\widehat{\tilde{\phi}}}(y)\right)^2\right]
\nonumber\\&&\hspace{1cm}-
\f{\epsilon(g_2(t)-g_1(t))}{L_2-L_1}\int_{\R}dy
\left(\partial_y{\widehat{\tilde{\phi}}}(y)\right)^2
\nonumber\\&&\hspace{1cm}+ \frac{\epsilon}{2}\left[\sum_{j=1,2}
\int_{\R}dy\f{(-1)^j\dot{g}_j(t)\widehat{\tilde{\xi}}(y)}{
L_2-L_1}\left(
\partial_y\widehat{\tilde{\phi}}(y)(y-\bar{L}_{j})
+\f{1}{2}\widehat{\tilde{\phi}}(y)\right)+hc\right]+
{\mathcal O}(\epsilon^2).
\end{eqnarray}
In this case, the part of the Hamiltonian that describes the
interaction between the field and the mirrors is also dependent on
$\epsilon$. However, in general, it is impossible to adequately
describe this part. For that reason,  the reactive part of the
motion force can seldom be calculated.

For instance, if we consider the generalization to the Lagrangian
density $(\ref{lbc})$, i.e.
\begin{eqnarray}
\f{1}{2}\left((\partial_t\phi)^2-(\partial_x\phi)^2\right)
-\sum_{j=1,2}\alpha_j\phi^2 \delta(x-L_j(t;\epsilon)),
\end{eqnarray}
then, the part of the quantum Hamiltonian that describes the
interaction is
\begin{eqnarray}
\sum_{j=1,2}\alpha_j\hat{\phi}^2(t,L_j(t;\epsilon))
=\sum_{j=1,2}\alpha_j\left(\widehat{\tilde{\phi}}(L_j)\right)^2
- \f{\epsilon(g_2(t)-g_1(t))}{L_2-L_1}\sum_{j=1,2}\alpha_j
\left(\widehat{\tilde{\phi}}(L_j)\right)^2+{\mathcal O}(\epsilon^2).
\end{eqnarray}
And, since
$$\f{1}{2}\int_{\R}dy
\left[\left(\widehat{\tilde{\xi}}(y)\right)^2+
\left(\partial_y{\widehat{\tilde{\phi}}}(y)\right)^2\right]+
\sum_{j=1,2}\alpha_j\left(\widehat{\tilde{\phi}}(L_j)\right)^2$$
is the  Hamiltonian of the system when the two mirrors are at rest,
that is the ``free'' Hamiltonian, we can conclude that, in the
interaction picture, while the mirrors move the full Hamiltonian of
the system is given by
\begin{eqnarray}&&
\hat{H}_I(t)=- \f{\epsilon(g_2(t)-g_1(t))}{L_2-L_1}\left[\int_{\R}dy
\left(\partial_y{\widehat{\tilde{\phi}}}_I(y)\right)^2
+\sum_{j=1,2}\alpha_j
\left(\widehat{\tilde{\phi}}_I(L_j)\right)^2\right]
\nonumber\\&&\hspace{1cm}+ \f{\epsilon}{2}\left[\sum_{j=1,2}
\int_{\R}dy\f{(-1)^j\dot{g}_j(t)\widehat{\tilde{\xi}}_I(y)}{
L_2-L_1}\left(
\partial_y\widehat{\tilde{\phi}}_I(y)(y-\bar{L}_{j})
+\f{1}{2}\widehat{\tilde{\phi}}_I(y)\right)+hc\right]+
{\mathcal O}(\epsilon^2).
\end{eqnarray}

Finally, we prove that our dissipative part of the motion force
coincides with the one obtained in \cite{jr9236}. For times $\tau$
larger than the stopping time, our quantum evolution operator,
in the linear approximation, is
${\mathcal T}^{\tau}=Id.-i\int_{\R}dt\, \hat{H}_I(t)$. We are
interested in the term
\begin{eqnarray}
A_1\equiv \f{\epsilon}{2}\left[\sum_{j=1,2}
\int_{\R}dy\f{(-1)^j\dot{g}_j(t)\widehat{\tilde{\xi}}_I(y)}{
L_2-L_1}\left(
\partial_y\widehat{\tilde{\phi}}_I(y)(y-\bar{L}_{j})
+\f{1}{2}\widehat{\tilde{\phi}}_I(y)\right)+hc\right].
\end{eqnarray}
Integrating by parts, and using the fact that
$\widehat{\tilde{\phi}}_I$ and $\widehat{\tilde{\xi}}_I$ are free
fields, it follows that
\begin{eqnarray}&&
A_1\equiv -\epsilon\sum_{j=1,2}\int_{\R}dt g_j(t)\hat{F}_j(t)
+\epsilon\int_{\R}dt \int_{\R}dy\f{{g}_2(t)-g_1(t)}{
L_2-L_1}\left(\partial_y{\widehat{\tilde{\phi}}}_I(y)\right)^2
\nonumber\\&& -\f{\epsilon}{2}\lim_{\delta\rightarrow 0}\int_{\R}dt
\f{{g}_2(t)-g_1(t)}{
L_2-L_1}\sum_{j=1,2}\widehat{\tilde{\phi}}_I(L_j)
\left(\partial_y\widehat{\tilde{\phi}}_I(L_j-|\delta|)-
\partial_y\widehat{\tilde{\phi}}_I(L_j+|\delta|)\right).
\end{eqnarray}
Now, from expression (2.4) in Ref.~\cite{bc95},
$$\lim_{\delta\rightarrow 0}
\left(\partial_y\widehat{\tilde{\phi}}_I(L_j+|\delta|)
-
\partial_y\widehat{\tilde{\phi}}_I(L_j-|\delta|)\right)
=2\alpha\widehat{\tilde{\phi}}_I(L_j),$$ we obtain that
\begin{eqnarray}&&
A_1\equiv -\epsilon\sum_{j=1,2}\int_{\R}dt g_j(t)\hat{F}_j(t)
+\epsilon\int_{\R}dt \int_{\R}dy\f{{g}_2(t)-g_1(t)}{
L_2-L_1}\left(\partial_y{\widehat{\tilde{\phi}}}_I(y)\right)^2
\nonumber\\&&\hspace*{15mm} +{\epsilon}\int_{\R}dt
\f{{g}_2(t)-g_1(t)}{
L_2-L_1}\sum_{j=1,2}\alpha_j\left(\widehat{\tilde{\phi}}_I(L_j)
\right)^2.
\end{eqnarray}

And, finally, inserting this expression into $\int_{\R}
dt\hat{H}_I(t)$, we conclude that, for times $\tau$ beyond the
stopping time, we have
\begin{eqnarray}
{\mathcal T}^{\tau}=Id+i\epsilon\sum_{j=1,2}\int_{\R}dt
g_j(t)\hat{F}_j(t),
\end{eqnarray}
as we wanted to demonstrate.

\section{Discussion and conclusions}

By way of the consideration of physically plausible mirrors and
of a canonical use of the Hamiltonian approach, we have showed that the problem
of the negative energies that appears in the Davis-Fulling model can
be avoided, the main difference with respect to their approach (and other's)
being that partially transmitting mirrors, which are transparent to
sufficiently high frequencies ---thus providing a natural and physically sound
renormalization--- are here being considered.

We have also proved that our method provides a dissipative
radiative-reaction force that fully agrees with the dissipative
force derived in Refs.~\cite{jr92} and \cite{bc95}. On the other
hand, the motion force calculated using the Hamiltonian approach
contains some reactive terms, proportional to the
the mirrors's acceleration, which do not appear in the results obtained  in
Refs.~\cite{jr92} and \cite{jr9236}. Those terms are fundamental in
order to ensure that the energy remains positive at any time, and
consequently, to guarantee the validity of the concept of particle
also during the oscillation of the mirrors, which is certainly fast but,
in the proposed experimental settings, still very small as compared with the speed
of light \cite{kbo1}.

We have also seen explicitly that albeit a possible (and quite simple)
solution to this disagreement could be to perform a
mass renormalization that completely eliminates the reactive terms
proportional to the the mirror's acceleration (see Ref.~\cite{bc95}), it
would turn out in this case that the definition of particle itself
would be impossible to maintain at any time while the mirrors move,
what would be indeed remarkable, in view of the relatively small velocities
involved. Quite on the contrary, and in consonance with the realistic
boundary conditions imposed by us on the mirrors (which led precisely to
these additional terms), we take as the most reasonable and physically
meaningful renormalization condition to impose, to keep those terms
in full, in which case the definition of particle during the
oscillation of the mirrors can be consistently preserved, as well
as the fundamental laws of physics too, at any time $t$ during the process.
In plain words, as a bonus, the fundamental principle of energy conservation
holds during the whole evolution towards the end state.

To finish, we must mention that the dynamical Casimir effect is being discussed
right now by many groups in different contexts, and that the growing potential of this subject,
both as a fundamental phenomenon as well as for the number and importance of its
applications, is out of question. At very large scales, it is being considered in
theoretical Cosmology as a most natural explanation of the observed acceleration
in the Universe expansion (termed as dark energy) \cite{eecos}.
And in a very different context, some laboratory experiments have recently been proposed which
would provide an extremely nice, alternative proof of the validity of General Relativity
and to some semiclassical approaches to Quantum Gravity. In addition, they open the way to practical applications of the Casimir effect in nano-electronics and other technologies. In those contexts some recent papers have appeared which deserve careful consideration (see e.g. \cite{c2,kbo1}).


\section*{Acknowledgments}
This work has been supported in part by MEC (Spain), projects
MTM2005-07660-C02-01, BFM2006-02842, PR2006-0145, FIS2005-25313-E,
and by AGAUR (Gene\-ra\-litat de Catalunya), contract 2005SGR-00790.


\begin{thebibliography}{99}

\bibitem{dk96} V.V. Dodonov and A.B. Klimov, Phys. Rev. {\bf A53}, 2664 (1996).

\bibitem{jjps97} J.Y. Ji, H.H. Jung, J.W. Park, and K.S. Soh, Phys. Rev. {\bf A56}, 4440
(1997).

\bibitem{cdm02} M. Crocce, D.A.R. Dalvit, F.D. Mazzitelli,  Phys. Rev. {\bf
A66},
033811 (2002).

\bibitem{sps02} R. Sch\"utzhold, G. Plunien, and G. Soff,
  Phys. Rev. {\bf  A65}, 043820 (2002); G. Schaller, R. Sch\"utzhold,
  G. Plunien, and G. Soff, Phys. Rev. {\bf  A66}, 023812 (2002).

\bibitem{ru05} M.  Ruser, J. Opt. {\bf  B7}, 100 (2005).

\bibitem{bmoo} I.~Brevik, K.~Milton, S.D.~Odintsov and K.~Osetrin,
Phys.~Rev.~{\bf D62}, 064005 (2000).

\bibitem{fd76} S.A. Fulling and P.C.W. Davies, Proc. Roy. Soc. Lond.
{\bf  A348}, 393 (1976).

\bibitem{he2} J. Haro and E. Elizalde, Phys. Rev. Lett. {\bf 97}, 130401 (2006).

\bibitem{fv82} L.H. Ford and A. Vilenkin, Phys. Rev. {\bf D25}, 2569
(1982).

\bibitem{be94} G. Barton and C. Eberlein,
Ann. Phys. (NY) {\bf 227}, 222 (1994); R.  G\"utig and C. Eberlein,
 J. Phys. {\bf  A31}, 6819 (1998).

\bibitem{e93} C. Eberlein, J. Phys.  {\bf  I3}, 2151 (1993).

\bibitem{bc95} G. Barton and A. Calogeracos, Ann. Phys. (NY)
{\bf  238}, 227 (1995); A. Calogeracos and G. Barton,
 Ann. Phys. (NY) {\bf  238}, 268 (1995).

\bibitem{jr9236} M.-T. Jaekel and S. Reynaud,   J. Phys.
{\bf  I2}, 149 (1992);  {\bf I3} 1093 (1993); A. Lambrecht, M.-T.
Jaekel, and S. Reynaud, Phys. Rev. Lett. {\bf 77}, 615 (1996).

\bibitem{kbo1} W.-J. Kim, J.H. Brownell, and R. Onofrio, Phys. Rev.
Lett. {\bf 96}, 200402 (2006).

\bibitem{rt85} M.  Razavy and J. Terning, Phys. Rev.
{\bf  D31}, 307 (1985).

\bibitem{l94} C.K. Law, Phys. Rev. {\bf  A49}, 433 (1994);
{\bf  A51}, 2537 (1995).

\bibitem{sps98} R.  Sch\"utzhold, G. Plunien and G. Soff,
 Phys. Rev.  {\bf  A57}, 2311 (1998).

\bibitem{js96} H.  Johnston and S. Sarkar, J. Phys.
{\bf  A29}, 1741 (1996).

\bibitem{haro1} J. Haro, Int. Jour. Theor. Phys. 46, 951 (2007); 46, 1003 (2007).

\bibitem{jr91} M.-T. Jaekel and S. Reynaud,  J. Phys.
{\bf  I1}, 1395 (1991).

\bibitem{gmm94} A.A. Grib, S.G. Mamayev, and V.M. Mostepanenko,
{\it Vacuum Quantum Effects in Strong Fields} (Friedman Laboratory
Publishing, 1994).

\bibitem{mo70} G.T.  Moore, J. Math. Phys. {\bf 11}, 2679 (1970).

\bibitem{op01} N. Obadia  and R. Parentani,
Phys. Rev. {\bf  D64}, 044019 (2001).

\bibitem{bd82} N.D. Birrell and C.P.W. Davies, {\it  Quantum Fields
in Curved Space} (Cambridge University Press, 1982).

\bibitem{eea} E. Elizalde, J. Phys. {\bf A34}, 3025 (2001);
E. Elizalde, M. Bordag, and K. Kirsten, J. Phys. {\bf A31}, 1743 (1998);
M. Bordag, E. Elizalde, and K. Kirsten, J. Math. Phys. {\bf 37}, 895 (1996);
M. Bordag, E. Elizalde, K. Kirsten, and S. Leseduarte, Phys. Rev. {\bf D56},
4896 (1997); E. Elizalde, L. Vanzo, and  S. Zerbini, Commun. Math. Phys.
{\bf 194}, 613 (1998); E. Elizalde, Commun. Math. Phys. {\bf 198}, 83 (1998);
Nuovo Cim. {\bf 104B}, 685 (1989); J. Phys. {\bf A22}, 931 (1989); E. Elizalde
and A. Romeo, Phys. Rev. {\bf D40}, 436 (1989).

\bibitem{jaffeea} R.L. Jaffe, Phys. Rev. {\bf D72}, 021301 (2005); T. Emig, R.L. Jaffe,
M. Kardar, and A. Scardicchio, Phys. Rev. Lett. {\bf 96}, 080403 (2006); N. Graham,
R.L. Jaffe, and H. Weigel, Int. J. Mod. Phys. {\bf A17}, 846 (2002).

\bibitem{jr92} M.-T. Jaekel and S. Reynaud, Quant. Opt. {\bf 4}, 39 (1992).

\bibitem{sb95} G.M. Salomone  and G. Barton, Phys. Rev. {\bf  A51}, 3506 (1995).

\bibitem{dmn1} D.A.R.  Dalvit and P.A. Maia Neto, Phys. Rev. Lett. {\bf 84}, 798 (2000);
P.A. Maia Neto and D.A.R.  Dalvit, Phys. Rev. {\bf  A62}, 042103
(2000).

\bibitem{eecos} E. Elizalde, S. Nojiri, S.D. Odintsov, and P. Wang, Phys. Rev. {\bf D71},
103504  (2005); E. Elizalde, S. Nojiri, S.D. Odintsov, and S. Ogushi, Phys. Rev. {\bf D67},
063515 (2003); E. Elizalde, J. Phys. {\bf A36}, L567 (2003); Phys. Lett. {\bf B516}, 143
(2001); J. Math. Phys. {\bf 35}, 6100 (1994); {\bf 35}, 3308 (1994);

\bibitem{c2} F. Capasso, Fellow, J.N. Munday, D. Iannuzzi, and H.B. Chan, IEEE J. Sel.
Topics Q. Electr. {\bf 13}, 400 (2007).


\end{thebibliography}
\end{document}